%% file: main.tex
\documentclass[conference]{IEEEtran}
\IEEEoverridecommandlockouts
\usepackage{cite}
\usepackage{amsmath,amssymb,amsfonts}
\usepackage{graphicx}
\usepackage{textcomp}

\usepackage{booktabs} % For formal tables
\usepackage{arydshln}
\usepackage{subcaption,verbatim,engord,enumitem,multirow,bm,caption}
\usepackage{algorithm,algorithmic}
\usepackage{amscd}
\usepackage{mathtools}
\usepackage{balance}

\newtheorem{definition}{Definition}
\newtheorem{proposition}{Proposition}
\newtheorem{theorem}{Theorem}
\newtheorem{proof}{Proof}

\usepackage{xcolor}
\def\BibTeX{{\rm B\kern-.05em{\sc i\kern-.025em b}\kern-.08em
    T\kern-.1667em\lower.7ex\hbox{E}\kern-.125emX}}
\begin{document}

\title{Scalable Realistic Recommendation Datasets through Fractal Expansions\\
}

\author{\IEEEauthorblockN{Francois Belletti}
\IEEEauthorblockA{\textit{Google AI} \\
\textit{Google}\\
belletti@google.com}
\and
\IEEEauthorblockN{Karthik Lakshmanan}
\IEEEauthorblockA{\textit{Google AI} \\
\textit{Google}\\
lakshmanan@google.com
}
\and
\IEEEauthorblockN{Walid Krichene}
\IEEEauthorblockA{\textit{Google AI} \\
\textit{Google}\\
walidk@google.com
}
\and
\IEEEauthorblockN{Yi-Fan Chen}
\IEEEauthorblockA{\textit{Google AI} \\
\textit{Google}\\
yifanchen@google.com
}
\and
\IEEEauthorblockN{John Anderson}
\IEEEauthorblockA{\textit{Google AI} \\
\textit{Google}\\
janders@google.com
}
}

\maketitle

\begin{abstract}
Recommender System research suffers currently from a disconnect between the size of academic data sets and the scale of industrial production systems.
In order to bridge that gap we propose to generate more massive user/item interaction data sets by expanding pre-existing public data sets.

User/item incidence matrices record interactions between users and items on a given platform as a large sparse matrix whose rows correspond to users and whose columns correspond to items.
Our technique expands such matrices to larger numbers of rows (users), columns (items) and non zero values (interactions) while preserving key higher order statistical properties.

We adapt the Kronecker Graph Theory to user/item incidence matrices and show that the corresponding \emph{fractal expansions} preserve the fat-tailed distributions of user engagements, item popularity and singular value spectra of user/item interaction matrices.
Preserving such properties is key to building large realistic synthetic data sets which in turn can be employed reliably to benchmark Recommender Systems and the systems employed to train them.

We provide algorithms to produce such expansions and apply them to the MovieLens 20 million data set comprising 20 million ratings of 27K movies by 138K users.
The resulting expanded data set has 10 billion ratings, 864K items and 2 million users in its smaller version and can be scaled up or down. A larger version features 655 billion ratings, 7 million items and 17 million users.
\end{abstract}

\begin{IEEEkeywords}
Machine Learning, Deep Learning, Recommender Systems, Graph Theory, Simulation
\end{IEEEkeywords}

\section{Introduction}
\input{introduction}

\section{Related work}
\input{related_work}

\section{Fractal expansions of user/item interaction data sets}
\input{fractal_expansions}

\section{Statistical properties of Kronecker fractal expansions}\label{sec:theory}
\input{theoretical_properties}

\section{Experimentation on MovieLens 20 million data}
\input{experiments}

\section{Conclusion}
\input{conclusion}

\bibliographystyle{acm}
{\small
\bibliography{biblio}
}
\newpage

\section*{Appendix}
\input{appendix}

\end{document}

%% file: introduction.tex
Machine Learning (ML) benchmarks compare the capabilities of models, distributed training systems and linear algebra accelerators on realistic problems at scale.
For these benchmarks to be effective, results need to be reproducible by many different groups which implies that publicly shared data sets need to be available. 

Unfortunately, while Recommendation Systems constitute a key industrial application of ML at scale, large public data sets recording user/item interactions on online platforms are not yet available.
For instance, although the Netflix data set \cite{bennett2007netflix} and the MovieLens data set \cite{harper2016movielens} are publicly available, they are orders of magnitude smaller than proprietary data~\cite{covington2016deep,belletti2018factorized,zhao2018categorical}.

\begin{table}[t!]
\center
\caption{Size of MovieLens 20M~\cite{harper2016movielens} vs industrial dataset in~\cite{zhao2018categorical}.}\label{tb:dataset}
\setlength{\tabcolsep}{2.5pt}
\begin{tabular}{lcc}
\toprule
 & \textbf{MovieLens 20M} & \textbf{Industrial}\\
\midrule
\#users  & 138K & Hundreds of Millions\\
\midrule
\#items & 27K & 2M\\
\midrule
\#topics & 19 & 600K\\
\midrule
\#observations & 20M & Hundreds of Billions\\
\bottomrule
\end{tabular}
\end{table}

\textbf{Proprietary data sets and privacy:}
While releasing large anonymized proprietary recommendation data sets may seem an acceptable solution from a technical standpoint, it is a non-trivial problem to preserve user privacy while still maintaining useful characteristics of the dataset.
For instance,\cite{narayanan2006break} shows a privacy breach of the Netflix prize dataset. More importantly, publishing anonymized industrial data sets runs counter to user expectations that their data may only be used in a restricted manner to improve the quality of their experience on the platform.

Therefore, we decide not to make user data more broadly available to preserve the privacy of users. We instead choose to produce synthetic yet realistic data sets whose scale is commensurable with that of our production problems while only consuming already publicly available data.

\textbf{Producing a realistic MovieLens 10 billion+ dataset:}
In this work, we focus on the MovieLens dataset which only entails movie ratings posted publicly by users of the MovieLens platform.
The MovieLens data set has now become a standard benchmark for academic research in Recommender Systems, \cite{tang2018caser,krichene2018efficient,he2017neural,liang2016factorization,tu2018structural,zheng2016neural,zhou2018deep,rudolph2016exponential,zhang2017enabling,abdollahpouri2017controlling,lu2016sparse,bhargava2017active,jawanpuria2018unified,nimishakavi2018dual} are only few of the many recent research articles relying on MovieLens whose latest version~\cite{harper2016movielens} has accrued more than $800$ citations according to Google Scholar.
Unfortunately, the data set comprises only few observed interactions and more importantly a very small catalogue of users and items --- when compared to industrial proprietary recommendation data.

\begin{figure}
    \centering
    \includegraphics[width=0.5\textwidth]{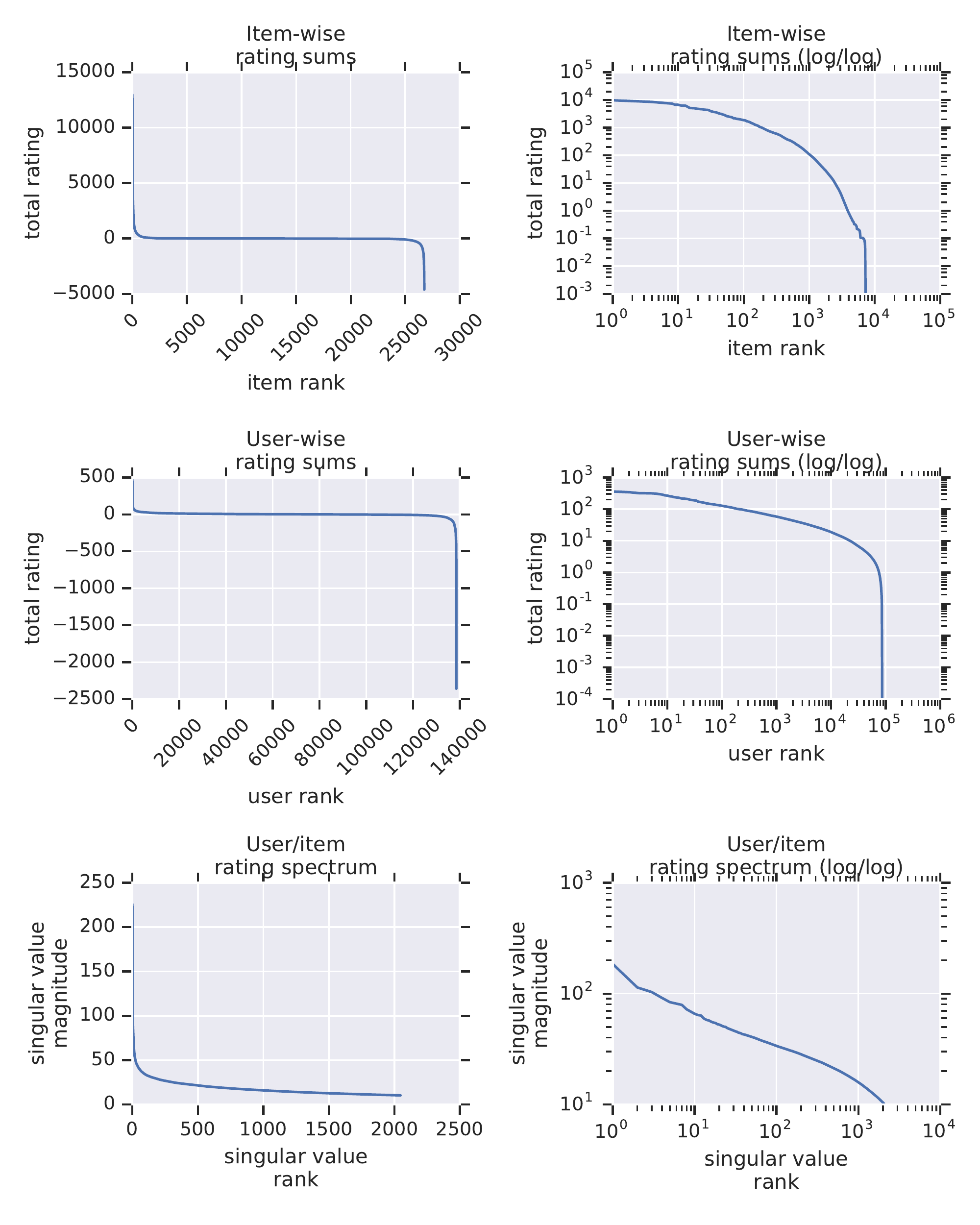}
        \caption{Key first and second order properties of the original MovieLens 20m user/item rating matrix (after centering and re-scaling into $[-1, 1]$) we aim to preserve while synthetically expanding the data set. \textbf{Top:} item popularity distribution (total ratings of each item). \textbf{Middle:} user engagement distribution (total ratings of each user). \textbf{Bottom:} dominant singular values of the rating matrix (core to the difficulty of matrix factorization tasks). In all log/log plots the small fraction of non-positive row-wise and column-wise sums are removed.}
    \label{fig:original_properties}
\end{figure}

In order to provide a new data set --- more aligned with the needs of production scale Recommender Systems ---
we aim at expanding publicly available data by creating a realistic surrogate.
The following constraints help create a production-size synthetic recommendation problem similar and at least as hard an ML problem as the original one for matrix factorization approaches to recommendations~\cite{koren2009matrix,he2017neural}:
\begin{itemize}
    \item orders of magnitude more users and items are present in the synthetic dataset;
    \item the synthetic dataset is realistic in that its first and second order statistics match those of the original dataset presented in Figure~\ref{fig:original_properties}.
\end{itemize}

Key first and second order statistics of interest we aim to preserve
are summarized in Figure~\ref{fig:original_properties} --- the details of their computation are given in Section~\ref{sec:theory}.

\textbf{Adapting Kronecker Graph expansions to user/item feedback:}
We employ the Kronecker Graph Theory introduced in~\cite{leskovec2005realistic} to achieve a suitable fractal expansion of recommendation data to benchmark linear and non-linear user/item factorization approaches for recommendations~\cite{koren2009matrix,he2017neural}.
Consider a recommendation problem comprising $m$ users and $n$ items.
Let $(R_{i,j})_{i=1 \dots m,j=1 \dots n}$ be the sparse matrix of recorded interactions (e.g. the rating left by the user $i$ for item $j$ if any and $0$ otherwise).
The key insight we develop in the present paper is that a carefully crafted fractal expansion of $R$ can preserve high level statistics of the original data set while scaling its size up by multiple orders of magnitudes.

Many different transforms can be applied to the matrix $R$ which can be considered a standard sparse 2 dimensional image. 
A recent approach to creating synthetic recommendation data sets consists in making parametric assumptions on user behavior by instantiating a user model interacting with an online platform~\cite{chaney2017algorithmic,schmit2017human}.
Unfortunately, such methods (even calibrated to reproduce empirical facts in actual data sets) do not provide strong guarantees that the resulting interaction data is similar to the original.
Therefore, instead of simulating recommendations in a parametric user-centric way as in~\cite{chaney2017algorithmic,schmit2017human}, we choose a non-parametric approach operating directly in the space of user/item affinity.
In order to synthesize a large realistic dataset in a principled manner, we adapt the Kronecker expansions which have previously been employed to produce large realistic graphs in~\cite{leskovec2005realistic}.
We employ a non-parametric analytically tractable simulation of the evolution of the user/item bi-partite graph to create a large synthetic data set.
Our choice is to trade-off realism for analytic tractability. We emphasize the latter.

While Kronecker Graphs Theory is developed in~\cite{leskovec2005realistic,leskovec2010kronecker} on square adjacency matrices, the Kronecker product operator is well defined on rectangular matrices and therefore we can apply a similar technique to user/item interaction data sets --- which was already noted in~\cite{leskovec2010kronecker} but not developed extensively.
The Kronecker Graph generation paradigm has to be changed with the present data set in other aspects however: we need to decrease the expansion rate to generate data sets with the scale we desire, not orders of magnitude too large. We need to do so while maintaining key conservation properties of the original algorithm~\cite{leskovec2010kronecker}.

In order to reliably employ Kronecker based fractal expansions on recommender system data we devise the following contributions:
\begin{itemize}
    \item we develop a new technique based on linear algebra to adapt fractal Kronecker expansions to recommendation problems;
    \item we demonstrate that key recommendation system specific properties of the original dataset are preserved by our technique;
    \item we also show that the resulting algorithm we develop is scalable and easily parallelizable as we employ it on the actual MovieLens 20 million dataset;
    \item we produce a synthetic yet realistic MovieLens 655 billion dataset to help recommender system research scale up in computational benchmark for model training.
\end{itemize}

The present article is organized as follows: we first recapitulate prior research on ML for recommendations and large synthetic dataset generation; we then develop an adaptation of Kronecker Graphs to user/item interaction matrices and prove key theoretical properties; finally we employ the resulting algorithm experimentally to MovieLens 20m data and validate its statistical properties.

%% file: related_work.tex
Recommender Systems constitute the workhorse of many e-commerce, social networking and entertainment platforms.
In the present paper we focus on the classical setting where the key role of a recommender system is to suggest relevant items to a given user.
Although other approaches are very popular such as content based recommendations~\cite{ricci2015recommender} or social recommendations~\cite{chaney2015probabilistic}, collaborative filtering remains a prevalent approach to the recommendation problem~\cite{linden2003amazon,sarwar2001item,resnick1994grouplens}.

\textbf{Collaborative filtering:}
The key insight behind collaborative filtering is to learn affinities between users and items based on previously collected user/item interaction data.
Collaborative filtering exists in different flavors.
Neighborhood methods group users by inter-user similarity and will recommend items to a given user that have been consumed by neighbors~\cite{ricci2015recommender}.
Latent factor methods such as matrix factorization~\cite{koren2009matrix} try to decompose user/item affinity as the result of the interaction of a few underlying representative factors characterizing the user and the item.
Although other latent models have been developed and could be used to construct synthetic recommendation data sets (e.g. Principal Component Analysis~\cite{jolliffe2011principal} or Latent Dirichlet Allocation~\cite{blei2003latent}), we focus on insights derived from matrix factorization.

The matrix factorization approach represents the affinity $a_{i,j}$ between a user $i$ and an item $j$ with an inner product $x^T_i y_j$ where $x_i$ and $y_j$ are two vectors in $\mathbb{R}^d$ representing the user and the item respectively.
Given a sparse matrix of user/item interactions $R=(r_{i,j})_{i = 1 \dots m, j=1 \dots n}$, user and item factors can therefore be learned by approximating $R$ with a low rank matrix $X Y^T$ where $X \in \mathbb{R}^{m, k}$ entails the user factors and $Y \in \mathbb{R}^{n, k}$ contains the item factors.
The data set $R$ represents ratings as in the MovieLens dataset~\cite{harper2016movielens} or item consumption ($r_{i,j}=1$ if and only if the user $i$ has consumed item $j$~\cite{bennett2007netflix}).
The matrix factorization approach is an example of a solution to the rating matrix completion problem which aims at predicting the rating of an item $j$ by a user $i$ which has not been observed yet and corresponds to a value of $0$ in the sparse original rating matrix.
Such a factorization method learns an approximation of the data that preserves a few higher order properties of the rating matrix $R$. 
In particular, the low rank approximation tries to mimic the singular value spectrum of the original data set.
We draw inspiration from matrix factorization to tackle synthetic data generation.
The present paper will adopt a similar approach to extend collaborative filtering data-sets. 
Besides trying to preserve the spectral properties of the original data,
we operate under the constraint of conserving its first and second order statistical properties.

\textbf{Deep Learning for Recommender Systems:}
Collaborative filtering has known many recent developments which motivate our objective of expanding public data sets in a realistic manner.
Deep Neural Networks (DNNs) are now becoming common in both non-linear matrix factorization tasks~\cite{wang2015collaborative,he2017neural,covington2016deep} and sequential recommendations~\cite{zhou2004intelligent,shani2005mdp,hariri2012context}.
%The latter rely typically on convolutional neural networks, attention networks or recurrent neural networks to map a sequence of temporally ordered user/item interactions to a set of related items~\cite{tang2018caser,li2017neural,wu2017rnn,beutel2018latent}.
The mapping between user/item pairs and ratings is generally learned by training the neural model to predict user behavior on a large data set of previously observed user/item interactions.

DNNs consume large quantities of data and are computationally expensive to train, therefore they give rise to commonly shared benchmarks aimed at speeding up training.
For training, a Stochastic Gradient Descent method is employed~\cite{lecun2015deep}
which requires forward model computation and back-propagation to be run on many mini-batches of (user, item, score) examples.
The matrix completion task still consists in predicting a rating for the interaction of user $i$ and item $j$ although $(i,j)$ has not been observed in the original data-set.
The model is typically run on billions of examples as the training procedure iterates over the training data set.

Model freshness is generally critical to industrial recommendations~\cite{covington2016deep} which implies that only limited time is available to re-train the model on newly available data.
The throughput of the trainer is therefore crucial to providing more engaging recommendation experiences and presenting more novel items.
Unfortunately, public recommendation data sets are too small to provide training-time-to-accuracy benchmarks that can be realistically employed for industrial applications.
Too few different examples are available in MovieLens 20m for instance and the number of different available items is orders of magnitude too small. 
In many industrial settings, millions of items (e.g. products, videos, songs) have to be taken into account by recommendation models.
The recommendation model learns an embedding matrices of size $(N, d)$ where $d \sim 10 - 10^3$
and $N \sim 10^6 - 10^9$ are typical values.
As a consequence, the memory footprint of this matrix may dominate that of the rest of the model by several orders of magnitude.
During training, the latency and bandwidth of the access to such embedding matrices have a prominent influence on the final throughput in examples/second.
Such computational difficulties associated with learning large embedding matrices are worthwhile solving in benchmarks.
A higher throughput enables training models with more examples which enables better statistical regularization and architectural expressiveness.
The multi-billion interaction size of the data set used for training is also a major factor that affects modeling choices and infrastructure development in the industry.

Our aim is therefore to enable a comparison of modeling approaches, software engineering frameworks and hardware accelerators for ML in the context of industry scale recommendations.
In the present paper we focus on enabling a better evaluation of examples/sec throughput and training-time-to-accuracy for Neural Collaborative Filtering Approaches~\cite{he2017neural} and Matrix Factorization Approaches~\cite{koren2009matrix}.
A major issue with this approach is, as we mentioned, the size of publicly available collaborative filtering data sets which is orders of magnitude smaller than production grade data (see Table~\ref{tb:dataset}) and has mis-representative orders of magnitudes in terms of numbers of distinct users and items.
The present paper offers a first solution to this problem by providing a simple and tractable non-parametric fractal approach to scaling up public recommendation data sets by several orders of magnitude.
Such data sets will help build a first set of benchmarks for model training accelerators based on publicly available data.
We plan to publish the expanded data. However, MovieLens 20m is publicly available and our method can already be applied to recreate such an expanded data set.
Meta-data is also common in industrial applications~\cite{covington2016deep,belletti2018factorized,beutel2018latent} but we consider its expansion outside the scope of this first development.

%% file: fractal_expansions.tex
The present section delineates the insights orienting our design decisions when expanding public recommendation data sets.

\subsubsection{Self-similarity in user/item interactions}

\begin{figure}
    \centering
    \includegraphics[width=0.5\textwidth,trim={1cm 5cm 6cm 0}]{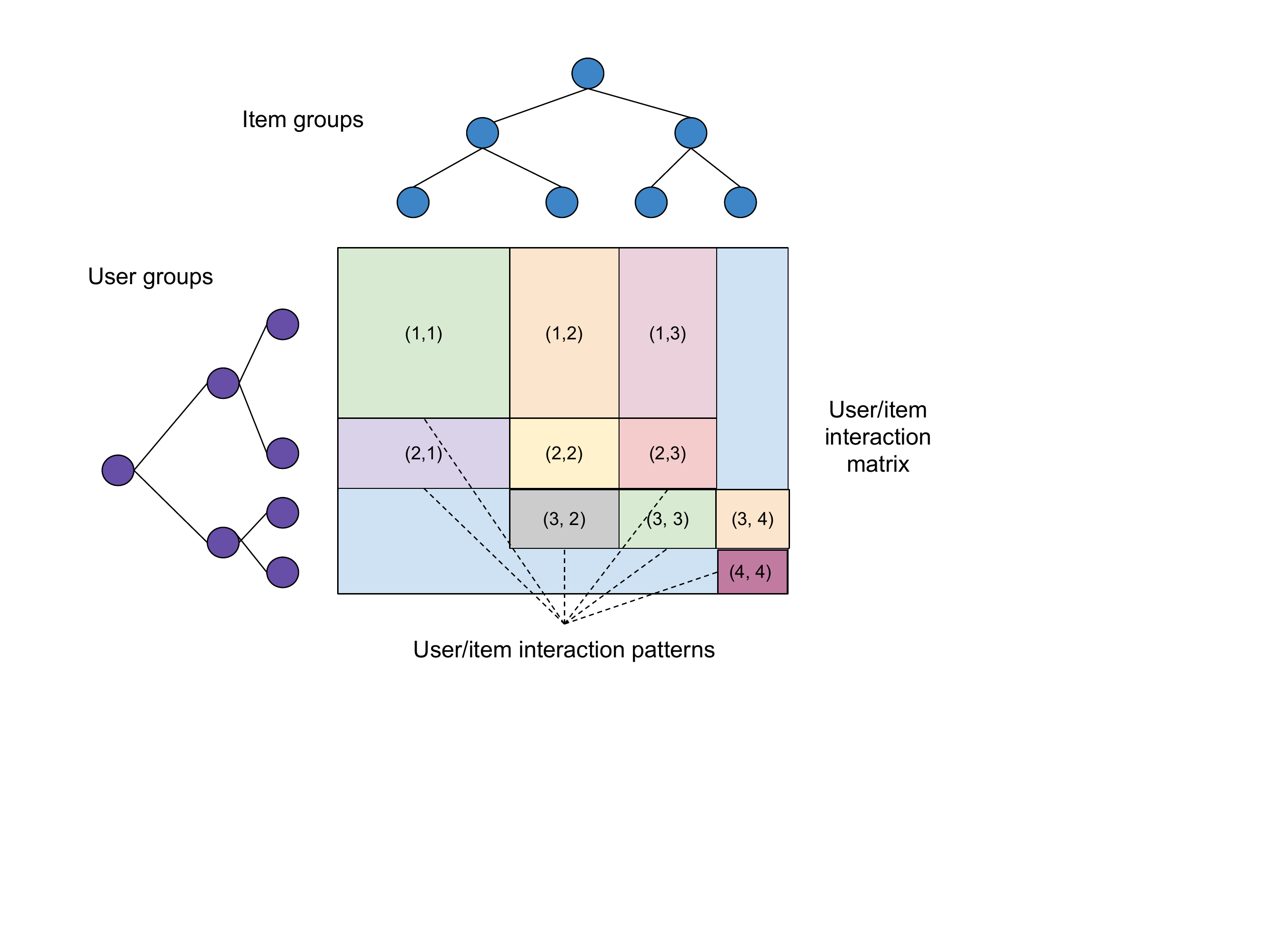}
    \caption{Typical user/item interaction patterns in recommendation data sets. Self-similarity appears as a natural key feature of the hierarchical organization of users and items into groups of various granularity.}
    \label{fig:user_item_patterns}
\end{figure}

Interactions between users and items follow a natural hierarchy in data sets where items can be organized in topics, genres, categories etc~\cite{zhao2018categorical}.
There is for instance an item-level fractal structure in MovieLens 20m with a tree-like structure of genres, sub-genres, and directors.
If users were clustered according to their demographics and tastes, another hierarchy would be formed~\cite{ricci2015recommender}.
The corresponding structured user/item interaction matrix is illustrated in Figure~\ref{fig:user_item_patterns}.
The hierarchical nature of user/item interactions (topical and demographic) makes the recommendation data set structurally self-similar (i.e. patterns that occur at more granular scales resemble those affecting coarser scales~\cite{mandelbrot1982fractal}).

One can therefore build a user-group/item-category incidence matrix with user-groups as rows and item-categories as columns
--- a coarse interaction matrix.
As each user group consists of many individuals and each item category comprises multiple movies, the original individual level user/item interaction matrix may be considered as an expanded version of the coarse interaction matrix.
We choose to expand the user/item interaction matrix by extrapolating this self-similar structure and simulating its growth to yet another level of granularity: each original item is treated as a synthetic topic in the expanded data set and each actual user is considered a fictional user group.

A key advantage of this fractal procedure is that it may be entirely non-parametric and designed to preserve high level properties of the original dataset.
In particular, a fractal expansion re-introduces the patterns originally observed in the entire real dataset within each block of local interactions of the synthetic user/item matrix.
By carefully designing the way such blocks are produced and laid out, we can therefore hope to produce a realistic yet much larger rating matrix.
In the following, we show how the Kronecker operator enables such a construction.

\subsubsection{Fractal expansion through Kronecker products}

The Kronecker product --- denoted $\otimes$ --- is a non-standard matrix operator with an intrinsic self-similar structure:
\begin{equation}
\label{eq:kronecker_product}
A \otimes B =
\begin{bmatrix}
    a_{11} B & \dots & a_{1n} B \\
    \vdots & \ddots & \vdots \\
    a_{m1} B & \dots & a_{mn} B
\end{bmatrix}
\end{equation}
where $A \in \mathbb{R}^{m,n}$, $B \in \mathbb{R}^{p,q}$ and $A \otimes B \in \mathbb{R}^{mp,nq}$.

In the original presentation of Kronecker Graph Theory~\cite{leskovec2005realistic} as well as the stochastic extension~\cite{mahdian2007stochastic} and the extended theory~\cite{leskovec2010kronecker}, the Kronecker product is the core operator enabling the synthesis of graphs with exponentially growing adjacency matrices.
As in the present work, the insight underlying the use of Kronecker Graph Theory in~\cite{leskovec2010kronecker} is to produce large synthetic yet realistic graphs.
The fractal nature of the Kronecker operator as it is applied multiple times (see Figure 2 in~\cite{leskovec2010kronecker} for an illustration) fits the self-similar statistical properties of real world graphs such as the internet, the web or social networks~\cite{leskovec2005realistic}.

If $A$ is the adjacency matrix of the original graph, fractal expansions are created in~\cite{leskovec2005realistic} by chaining Kronecker products as follows:
$$
    A \otimes A \dots A.
$$
As adjacency matrices are square, Kronecker Graphs are not employed on rectangular matrices in pre-existing work although the operation is well defined.
Another slight divergence between present and pre-existing work is that --- in their stochastic version --- Kronecker Graphs carry Bernouilli probability distribution parameters in $[0, 1]$ while MovieLens ratings are in $\left\{0.5, 1.0, \dots, 5.0\right\}$ originally and $[-1, 1]$ after we center and rescale them. 
We show that these differences do not prevent Kronecker products from preserving core properties of rating matrices.
A more important challenge is the size of the original matrix we deal with: $R \in \mathbb{R}^{(138 \times 10^3, 27 \times 10^3)}$.
A naive Kronecker expansion would therefore synthesize a rating matrix with $19$ billion users which is too large.

Thus, although Kronecker products seem like an ideal candidate for the mechanism at the core of the self-similar synthesis of a larger recommendation dataset, some modifications are needed to the algorithms developed in~\cite{leskovec2010kronecker}.

\subsubsection{Reduced Kronecker expansions}
We choose to synthesize a user/item rating matrix 
$$
\widetilde{R}
=
\widehat{R}
\otimes
R
$$
where
$
\widehat{R}
$
is a matrix derived from $R$ but much smaller (for instance $\widehat{R} \in \mathbb{R}^{128, 256}$).
For reasons that will become apparent as we explore some theoretical properties of Kronecker fractal expansions, we want to construct a smaller derived matrix $\widehat{R}$ that shares similarities with $R$. In particular, we seek $\widehat{R}$ with a similar row-wise sum distribution (user engagement distribution), column-wise distribution (item engagement distribution) and singular value spectrum (signal to noise ratio distribution in the matrix factorization).

\subsubsection{Implementation at scale and algorithmic extensions}
Computing a Kronecker product between two matrices $A$ and $B$ is an inherently parallel operation.
It is sufficient to broadcast $B$ to each element $(i, j)$ of $A$ and then multiply $B$ by $a_{i,j}$.
Such a property implies that scaling can be achieved.
Another advantage of the operator is that even a single machine can produce a large output data-set by sequentially iterating on the values of $A$. Only storage space commensurable with the size of the original matrix is needed to compute each block of the Kronecker product.
It is noteworthy that generalized fractal expansions can be defined by altering the standard Kronecker product. 
We consider such extensions here as candidates to engineer more challenging synthetic data sets. 
One drawback though is that these extensions may not preserve analytic tractability.

A first generalization defines a binary operator $\otimes_{F}$ with 
$
F
:
\mathbb{R} \times \mathbb{R}^{m,n} \times \mathbb{N}
\rightarrow
\mathbb{R}^{p, q}
$
as follows:
\begin{equation}
\label{eq:gen_kron_product}
A
\otimes_F
B= 
\begin{bmatrix}
    F(a_{11}, B, \omega_{11}) & \dots & F(a_{1n}, B, \omega_{1n}) \\
    \vdots & \ddots & \vdots \\
    F(a_{m1}, B, \omega_{m1}) & \dots & F(a_{mn}, B, \omega_{mn})
\end{bmatrix}    
\end{equation}

where $\omega_{11}, \dots, \omega_{mn}$ is a sequence of pseudo-random numbers.
Including randomization and non-linearity in $F$ appears as a simple way to synthesize data sets entailing more varied patterns. The algorithm we employ to compute Kronecker products is presented in Algorithm~\ref{alg:kron_exp}. The implementation we employ is trivially parallelizable. We only create a list of Kronecker blocks to dump entire rows (users) of the output matrix to file. This is not necessary and can be removed to enable as many processes to run simultaneously and independently as there are elements in $\widehat{R}$ (provided pseudo random numbers are generated in parallel in an appropriate manner).

\begin{algorithm}
\caption{Kronecker fractal expansion}
\label{alg:kron_exp}
\begin{algorithmic}
\FOR{$i = 1$ to $m$}
    \STATE{kBlocks $\gets $ empty list}
    \FOR{$j = 1$ to $n$}
        \STATE{$\omega$ $\gets$ next pseudo random number}
        \STATE{kBlock $\gets$ $F(\widehat{R}(i, j), R, w)$}
        \STATE{kBlocks append kBlock}
    \ENDFOR
    \STATE{outputToFile(kBlocks)}
\ENDFOR
\end{algorithmic}
\end{algorithm}

The only reason why we need a reduced version $\widehat{R}$ of $R$ is to control the size of the expansion.
Also, $A \otimes B$ and $B \otimes A$ are equal after a row-wise and a column-wise permutation.
Therefore, another family of appropriate extensions may be obtained by considering
\begin{equation}
\label{eq:random_sk_kron}
B \otimes_G A =
\begin{bmatrix}
    b_{11} G(A, \omega_{11}) & \dots & b_{1n} G(A, \omega_{1n}) \\
    \vdots & \ddots & \vdots \\
    b_{m1} G(A, \omega_{m1}) & \dots & b_{mn} G(A, \omega_{mn})
\end{bmatrix}
\end{equation}
where 
$
G
:
\mathbb{R}^{m,n} \times \mathbb{N}
\rightarrow
\mathbb{R}^{m',n'}
$
is a randomized sketching operation on the matrix $A$ which reduces its size by several orders of magnitude.
A trivial scheme consists in sampling a small number of rows and columns from $A$ at random. Other random projections~\cite{achlioptas2003database,li2006very,fradkin2003experiments} may of course be used.
The randomized procedures above produce a user/item interaction matrix where there is no longer a block-wise repetitive structure. Less obvious statistical patterns can give rise to more challenging synthetic large-scale collaborative filtering problems.

%% file: theoretical_properties.tex
After having introduced Kronecker products to self-similarly expand a recommendation dataset into a much larger one, we now demonstrate how the resulting synthetic user/item interaction matrix shares crucial common properties with the original.

\subsubsection{Salient empirical facts in MovieLens data}
First, we introduce the critical properties we want to preserve.
\emph{
Note that throughout the paper we present results on a centered version of MovieLens 20m. 
The average rating of $3.53$ is subtracted from all ratings (so that the $0$ elements of the sparse rating matrix match un-observed scores and not bad movie ratings).
Furthermore, we re-scale the centered ratings so that they are all in the interval $[-1, 1]$.
}
As a user/item interaction dataset on an online platform, one expects MovieLens to feature common properties of recommendation data sets such as ``power-law'' or fat-tailed distributions~\cite{zhao2018categorical}.

First important statistical properties for recommendations concern the distribution of interactions across users and across items. It is generally observed that such distributions exhibit a ``power-law'' behavior~\cite{zhao2018categorical,abdollahpouri2017controlling,oestreicher2012recommendation,cremonesi2010performance,park2008long,yin2012challenging,levy2010music,goel2010anatomy}.
To characterize such a behavior in the MovieLens data set, we take a look at the distribution of the total ratings along the item axis and the user axis. In other words, we compute row-wise and column-wise sums for the rating matrix $R$ and observe their distributions.
The corresponding ranked distributions are exposed in Figure~\ref{fig:original_properties} and do exhibit a clear ``power-law'' behavior for rather popular items. However we observe that tail items have a higher popularity decay rate. Similarly, the engagement decay rate increases for the group of less engaged users.

The other approximate ``power-law'' we find in Figure~\ref{fig:original_properties} lies in the singular value spectrum of the MovieLens dataset.
We compute the top $k$ singular values~\cite{horn1990matrix} of the MovieLens rating matrix $R$ by approximate iterative methods (e.g. power iteration) which can scale to its large $(138K, 27K)$ dimension.
The method yields the dominant singular values of $R$ and the corresponding singular vectors so that one can classically approximate $R$ by
$
    R \simeq U \Sigma V
$
where $\Sigma$ is diagonal of dimension $(k, k)$,
$U$ is column-orthogonal of dimension $(m, k)$ and $V$ is row-orthogonal of dimension $(k, n)$ ---
which yields the rank $k$ matrix closest to $R$ in Frobenius norm.

Examining the distribution of the $2048$ top singular values of $R$ in the MovieLens dataset (which has at most $27K$ non-zero singular values) in Figure~\ref{fig:original_properties} highlights a clear ``power-law'' behavior in the highest magnitude part of the spectrum of $R$.
We observe in the spectral distribution an inflection for smaller singular values whose magnitude decays at a higher rate than larger singular values.
Such a spectral distribution is as a key feature of the original dataset, in particular in that it conditions the difficulty of low-rank approximation approaches to the matrix completion problem.
Therefore, we also want the expanded dataset to exhibit a similar behavior in terms of spectral properties. 

In all the high level statistics we present, we want to preserve the approximate ``power-law'' decay as well as its inflection for smaller values. Our requirements for the expanding transform which we apply to $R$ are therefore threefold: \emph{we want to preserve the distributions of row-wise sums of $R$, column-wise sums of $R$ and singular value distribution of $R$}.
Additional requirements, beyond first and second order high level statistics will further increase the confidence in the realism of the expanded synthetic dataset.
Nevertheless, we consider that focusing on these first three properties is a good starting point.

It is noteworthy that we do not consider here the temporal structure of the MovieLens data set. We leave the study of sequential user behavior --- often found to be Long Range Dependent~\cite{pipiras2017long,belletti2017random,crane2008robust} --- and the extension of synthetic data generation to sequential recommendations~\cite{belletti2018factorized,tang2018caser,yu2016dynamic} for further work.

\subsubsection{Preserving MovieLens data properties while expanding it}
We now expose the fractal transform design we rely on to preserve the key statistical properties of the previous section.

\begin{definition}
Consider $A \in \mathbb{R}^{m, n} = (a_{i, j})_{i=1 \dots n, j=1 \dots m}$,
we denote the set $\left \{ \sum_{i = 1}^m a_{i, j}\right \}$ of row-wise sums of $A$  by $\mathcal{R}(M)$,
the set $\left \{ \sum_{j = 1}^n a_{i, j}\right \}$ of column-wise sums of $A$  by $\mathcal{C}(M)$,
and the set of non-zero singular values of $A$ by $\mathcal{S}(M)$.
\end{definition}

\begin{definition}
Consider an integer $i$ and a non-zero positive integer $p$,
we denote the integer part of $i - 1$ in base $p$
$
\lfloor{i - 1\rfloor}_p = \lfloor{\frac{i - 1}{p}\rfloor}
$
and the fractional part
$
\left\{i - 1\right\}_p = i - \lfloor{i - 1\rfloor}_p.
$
\end{definition}

First we focus on conservation properties in terms of row-wise and column-wise sums which correspond respectively to marginalized user engagement and item popularity distributions.
In the following, $\times$ denotes the Minkowski product of two sets, i.e. 
$A \times B = \left\{ a \times b \: | \: a \in A, b \in B \right\}$.

\begin{proposition}
\label{prop:c_r_distributions}
Consider $A \in \mathbb{R}^{m, n}$ and $B \in \mathbb{R}^{p, q}$ and
their Kronecker product $K = A \otimes B$. Then
$$\mathcal{R}(K) = \mathcal{R}(A) \times \mathcal{R}(B) \text{ and }
\mathcal{C}(K) = \mathcal{C}(A) \times \mathcal{C}(B).$$
\end{proposition}

\begin{proof}
Consider the $i^{th}$ row of $K$, by definition of $K$ the corresponding sum can be rewritten as follows: 
$
\sum_{j = 1 \dots n p} k_{i, j} 
=
\sum_{j = 1 \dots n p} 
a_{\lfloor{i - 1\rfloor}_p + 1, \lfloor{j - 1\rfloor}_q + 1} 
b_{\left\{i - 1\right\}_p, \left\{j - 1\right\}_q}$
which in turn equals
$$
\sum_{j = 1 \dots n}
\sum_{j'= 1 \dots q}
a_{\lfloor{i - 1\rfloor}_p + 1, j}
b_{\left\{i - 1\right\}_p, j'}.
$$
Refactoring the two sums concludes the proof for the row-wise sum properties. The proof for column-wise properties is identical.
$\blacksquare$
\end{proof}

\begin{theorem}
\label{th:spectral_distribution}
Consider $A \in \mathbb{R}^{m, n}$ and $B \in \mathbb{R}^{p, q}$ and
their Kronecker product $K = A \otimes B$. Then
$$\mathcal{S}(K) = \mathcal{S}(A) \times \mathcal{S}(B).$$
\end{theorem}
\begin{proof}
One can easily check that $(X Y) \otimes (V W) = (X \otimes V) (Y \otimes W)$ for any quadruple of matrices $X,Y,V,W$ for which the notation makes sense and that $(X \otimes Y)^T = X^T \otimes Y^T$.
Let $A = U_A \Sigma_A V_A$ be the SVD of $B$ and $B = V_B \Sigma_B V_B$ the SVD of $B$.
Then $(A \otimes B) = (U_A \otimes U_B) (\Sigma_A \otimes \Sigma_B) (V_A \otimes V_B)$.
Now, $(U_A \otimes U_B)^T (U_A \otimes U_B) = (U_A^T \otimes U_B^T) (U_A \otimes U_B) = (U_A^T U_A) \otimes (U_B^T U_B^T)$.
Writing the same decomposition for $(V_A \otimes V_B) (V_A^T \otimes V_B^T)$ and considering that $U_A$, $U_B$ are column-orthogonal while $V_A$, $V_B$ are row-orthogonal concludes the proof.
$\blacksquare$
\end{proof}

The properties above imply that knowing the row-wise sums, column-wise sums and singular value spectrum of the reduced rating matrix $\widehat{R}$ and the original rating matrix $R$ is enough to deduce the corresponding properties for the expanded rating matrix $\widetilde{R}$ --- analytically. As in~\cite{leskovec2010kronecker}, the Kronecker product enables analytic tractability while expanding data sets in a fractal manner to orders of magnitude more data.

\subsubsection{Constructing a reduced $\widehat{R}$ matrix with a similar spectrum}
Considering that the quasi ``power-law'' properties of $R$ imply --- as in~\cite{leskovec2010kronecker} --- that $\mathcal{S}(R) \times \mathcal{S}(R)$ has a similar distribution to $\mathcal{S}(R)$, we seek a small $\widehat{R}$ whose high order statistical properties are similar to those of $R$.
As we want to generate a dataset with several billion user/item interactions, millions of distinct users and millions of distinct items, we are looking for a matrix $\widehat{R}$ with a few hundred or thousand rows and columns.
The reduced matrix $\widehat{R}$ we seek is therefore orders of magnitude smaller than $R$.
% \textbf{First attempts through randomized sketching:} 
% Random projections~\cite{achlioptas2003database,li2006very,fradkin2003experiments} appear as valid candidates to reduce the size of $R$ down while preserving its spectral properties.
% Indeed, proper random projections aim at finding $\widehat{R}$ such that
% $\widehat{R} \widehat{R}^T \simeq R R^T$\
% Such methods unfortunately do not produce $\widehat{R}$ with similar properties to $R$ as the dimensions of $\widehat{R}$ are too small compared to $R$. For random projections, such an issue can be interpreted as a consequence of having few dimensions in the bounds given by the Johnson-Lindenstrauss lemma. For random hashing, too few hashing buckets lead to high levels of aliasing which in turn skew the spectrum of the reduced matrix $\widehat{R}$.
In order to produce a reduced matrix $\widehat{R}$ of dimensions $(1000, 1700)$ one could use the reduced size older MovieLens 100K dataset~\cite{harper2016movielens}.
Such a dataset can be interpreted as a sub-sampled reduced version of MovieLens 20m with similar properties. However the data sets have been collected seven years apart and therefore temporal non-stationarity issues become concerning. Also, we aim to produce an expansion method where the expansion multipliers can be chosen flexibly by practitioners. 
In our experiments, it is noteworthy that naive uniform user and item sampling strategies have not yielded smaller matrices $\widehat{R}$ with similar properties to $R$ in our experiments.
Different random projections~\cite{achlioptas2003database,li2006very,fradkin2003experiments} could more generally be employed however we rely on a procedure better tailored to our specific statistical requirements.

We now describe the technique we employed to produce a reduced size matrix $\widehat{R}$ with first and second order properties close to $R$ which in turn led to constructing an expansion matrix $\widetilde{R} = \widehat{R} \otimes R$ similar to $R$.
We want the dimensions of $\widehat{R}$ to be $(m',n')$ with $m' << m$ and $n' << n$.
Consider again the approximate Singular Value Decomposition (SVD)~\cite{horn1990matrix} of $R$ with the $k=min(m',n')$ principal singular values of $R$:
\begin{equation}
    \label{eq:R_svd}
    R \simeq U \Sigma V
\end{equation}
where $U \in \mathbb{R}^{n, k}$ has orthogonal columns, 
$V \in \mathbb{R}^{k, m}$ has orthogonal rows, 
and $\Sigma \in \mathbb{R}^{k, k}$ is diagonal with non-negative terms.

To reduce the number of rows and columns of $R$ while preserving its top $k$ singular values a trivial solution would consist in replacing $U$ and $V$ by a small random orthogonal matrices with few rows and columns respectively.
Unfortunately such a method would only seemingly preserve the spectral properties of $R$ as the principal singular vectors would be widely changed. Such properties are important: one of the key advantages of employing Kronecker products in~\cite{leskovec2010kronecker} is the preservation of the network values, i.e. the distributions of singular vector components of a Graph's adjacency matrix.

To obtain a matrix $\widetilde{U} \in \mathbb{R}^{n', k}$ with fewer rows than $U$ but column-orthogonal and similar to $U$ in the distribution of its values we use the following procedure.
We re-size $U$ down to $n'$ rows with $n' < n$ through an averaging-based down-scaling method that can classically be found in standard image processing libraries (e.g. skimage.transform.resize in the scikit-image library~\cite{van2014scikit}).
Let $\bar{U} \in \mathbb{R}^{n', k}$ be the corresponding resized version of $U$. 
We then construct $\widetilde{U}$ as the column orthogonal matrix in $\mathbb{R}^{n', k}$ closest in Frobenius norm to $\bar{U}$.
Therefore as in~\cite{golub2012matrix} we compute
\begin{equation}
    \label{eq:U_reduction}
    \widetilde{U} = \bar{U} \left(\bar{U}^T \bar{U}\right)^{-1/2}.  
\end{equation}
We apply a similar procedure to $V$ to reduce its number of columns which yields a row orthogonal matrix $\widetilde{V} \in \mathbb{R}^{k, m'}$ with $m' < m$.
The orthogonality of $\widetilde{U}$ (column-wise) and $\widetilde{V}$ (row-wise) guarantees that the singular value spectrum of
\begin{equation}
    \label{eq:R_hat_construction}
    \widehat{R} = \widetilde{U} \Sigma \widetilde{V}
\end{equation}
consists exactly of the $k=min(m',n')$ leading components of the singular value spectrum of $R$.
Like $R$, $\widehat{R}$ is re-scaled to take values in $[-1, 1]$.
The whole procedure to reduce $R$ down to $\widehat{R}$ is summarized in Algorithm~\ref{alg:reduced_matrix}.

\begin{algorithm}
\caption{Compute reduced matrix $\widehat{R}$}
\label{alg:reduced_matrix}
\begin{algorithmic}
\STATE{$(U, \Sigma, V) \gets$ sparseSVD($R,k$)}
\STATE{$\bar{U} \gets$ imageResize($U, n', k$)}
\STATE{$\bar{V} \gets$ imageResize($V, k, m'$)}
\STATE{$\widetilde{U} \gets \bar{U} \left(\bar{U}^T \bar{U}\right)^{-1/2}$}
\STATE{$\widetilde{V} \gets \left(\bar{V} \bar{V}^T\right)^{-1/2} \bar{V}$}
\STATE{$\widehat{R}_{temp} \gets \widetilde{U} \Sigma \widetilde{V}$}
\STATE{$M \gets max(\widehat{R}_{temp})$}
\STATE{$m \gets min(\widehat{R}_{temp})$}
\STATE{return $\widehat{R}_{temp} / (M - m)$}
\end{algorithmic}
\end{algorithm}

We verify empirically that the distributions of values of the reduced singular vectors in $\widetilde{U}$ and $\widetilde{V}$ are similar to those of $U$ and $V$ respectively to preserve first order properties of $R$ and value distributions of its singular vectors.
Such properties are demonstrated through numerical experiments in the next section.

%% file: experiments.tex
The MovieLens 20m data comprises 20m ratings given by $138$ thousand users to $27$ thousand items.
In the present section, we demonstrate how the fractal Kronecker expansion technique we devised and presented helps scale up this dataset to orders of magnitude more users, items and interactions --- all in a parallelizable and analytically tractable manner.

\subsubsection{Size of expanded data set}

In present experiments we construct a reduced rating matrix $\widehat{R}$ of size $(16, 32)$ which implies the resulting expanded data set will comprise $10$ billion interactions between $2$ million users and $864$K items.
In appendix, we present the results obtained for a reduced rating matrix $\widehat{R}$ of size $(128, 256)$ and a synthetic data set consisting of $655$ billion interactions between $17$ million users and $7$ million items.

Such a high number of interactions and items enable the training of deep neural collaborative models such as the Neural Collaborative Filtering model~\cite{he2017neural} with a scale which is now more representative of industrial settings.
Moreover, the increased data set size helps construct benchmarks for deep learning software packages and ML accelerators that employ the same orders of magnitude than production settings in terms of user base size, item vocabulary size and number of observations.

\subsubsection{Empirical properties of reduced $\widehat{R}$ matrix}

The construction technique of $\widehat{R}$ had for objective to produce, just like in~\cite{leskovec2010kronecker}, a matrix sharing the properties of $R \otimes R$ though smaller in size.
To that end, we aimed at constructing a matrix $\widehat{R}$ of dimension $(16, 32)$ with properties close to those of $R$ in terms of column-wise sum, row-wise sum and singular value spectrum distributions.

We now check that the construction procedure we devised does produce a $\widehat{R}$ with the properties we expected.
As the impact of the re-sizing step is unclear from an analytic stand-point, we had to resort to numerical experiments to validate our method.

\begin{figure}
    \centering
    \includegraphics[width=\linewidth]{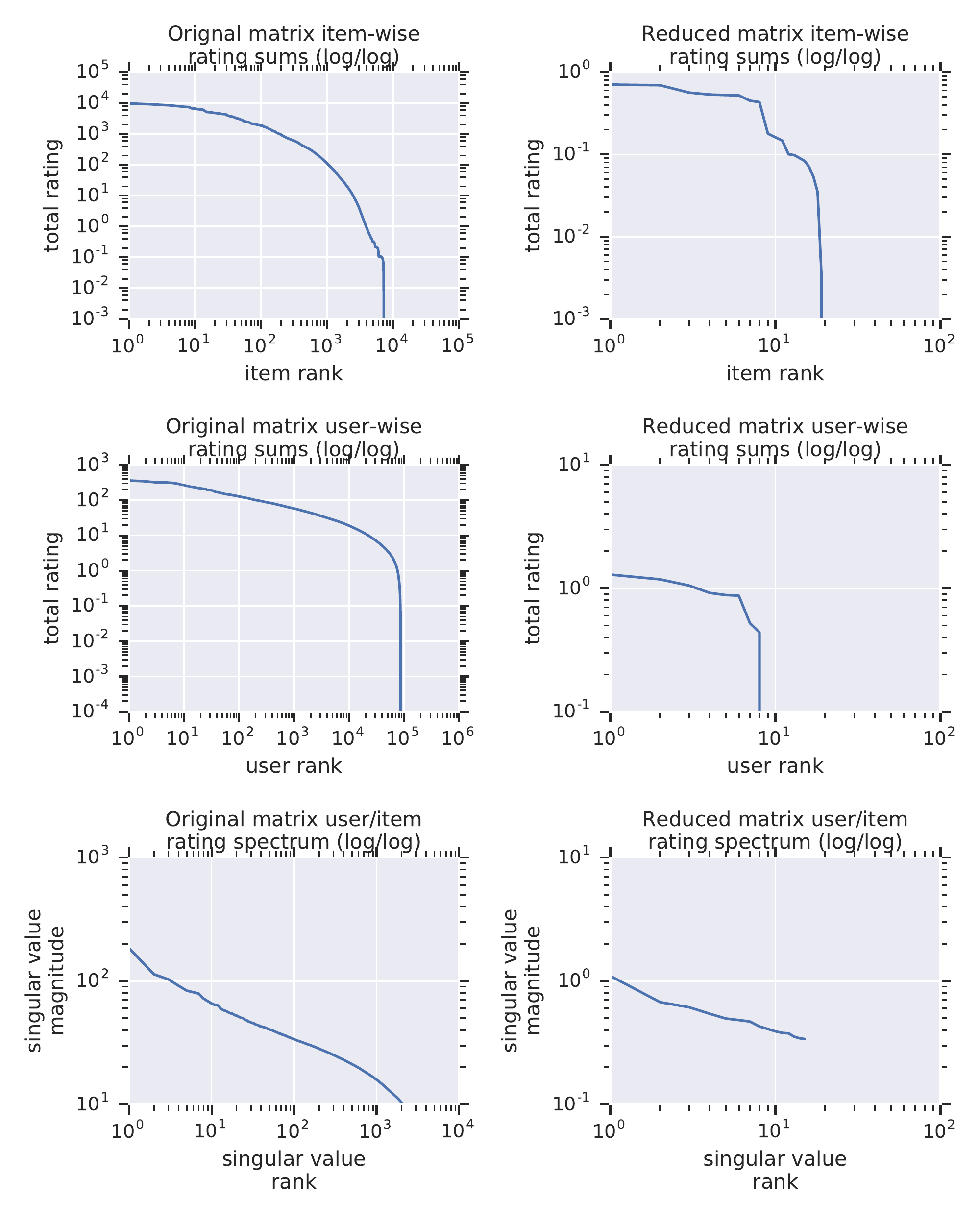}
    \caption{Properties of the reduced dataset $\widehat{R} \in \mathbb{R}^{16, 32}$ built according to steps~\ref{eq:R_svd}, \ref{eq:U_reduction} and \ref{eq:R_hat_construction}. We validate the construction method numerically by checking that the distribution of row-wise sums, column-wise sums and singular values are similar between $R$ and $\widehat{R}$. Note here that as $R$ is large we only compute its leading singular values. As we want to preserve statistical ``power-laws'', we focus on preservation of the relative distribution of values and not their magnitude in absolute.}
    \label{fig:small_properties}
\end{figure}

In Figure~\ref{fig:small_properties}, one can assess that the first and second order properties of $R$ and $\widehat{R}$ match with high enough fidelity. In particular, the higher magnitude column-wise and row-wise sum distributions follow a ``power-law'' behavior similar to that of the original matrix. Similar observations can be made about the singular value spectra of $\widehat{R}$ and $R$.

There is therefore now a reasonable likelihood that our adapted Kronecker expansion --- although somewhat differing from the method originally presented in~\cite{leskovec2010kronecker} --- will enjoy the same benefits in terms of enabling data set expansion while preserving high order statistical properties.

\subsubsection{Empirical properties of the expanded data set $\widetilde{R}$}
We now verify empirically that the expanded rating matrix $\widetilde{R} = \widehat{R} \otimes R$ does share common first and second order properties with the original rating matrix $R$. The new data size is $2$ orders of magnitude larger in terms of number of rows and columns and $4$ orders of magnitude larger in terms of number of non-zero terms.
Notice here that, as in general $\widehat{R}$ is a dense matrix, the level of sparsity of the expanded data set is the same as that of the original.

Another benefit of using a fractal expansion method with analytic tractability, is that we can deduce high order statistics of the expanded data set beforehand without having to instantiate it.
In particular, Proposition~\ref{prop:c_r_distributions} implies that knowing the column-wise and row-wise sum distributions of $\widehat{R}$ and $R$ is sufficient to determine the corresponding marginals for the expanded data set $\widetilde{R} = \widehat{R} \otimes R$.
Similarly, the leading singular values of the Kronecker product can be computed with Theorem~\ref{th:spectral_distribution} just based on the leading singular values of $R$ and the singular values of $\widehat{R}$.

\begin{figure}
    \centering
    \includegraphics[width=\linewidth]{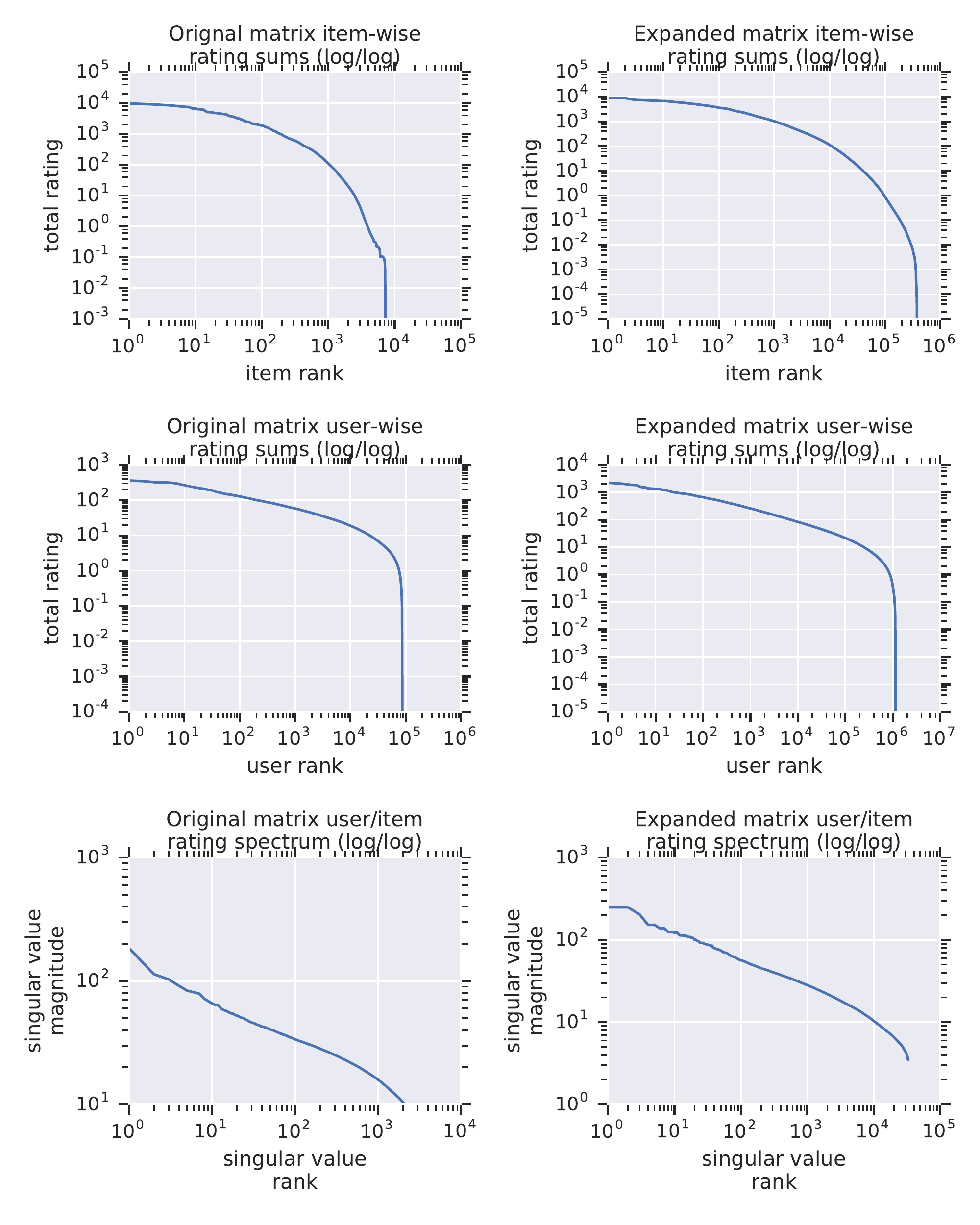}
    \caption{High order statistical properties of the expanded dataset $\widehat{R} \otimes R$. We validate the construction method numerically by checking that the distributions of row-wise sums, column-wise sums and singular values are similar between $R$ and $\widehat{R} \otimes R$. Here we leverage the tractability of Kronecker products as they impact column-wise and row-wise sum distributions as well as singular value spectra. The plots corresponding to the extended dataset are derived analytically based on the corresponding properties of the reduced matrix $\widehat{R}$ and the original matrix $R$.
    Note here that as we only computed the leading singular values of $R$, we only show the leading singular values of $\widehat{R} \otimes R$.
    In all plots we can observe the preservation of the linear log-log correspondence for the higher values in the distributions of interest (row-wise sums, column-wise sums and singular values) as well as the accelerated decay of the smaller values in those distributions.
    }
    \label{fig:expanded_properties}
\end{figure}

In Figure~\ref{fig:expanded_properties}, one can confirm that the spectral properties of the expanded data set as well as the user engagement (row-wise sums) and item popularity (column-wise sums) are similar to those of the original data set.
Such observations indicate that the resulting data set is representative --- in its fat-tailed data distribution and quasi ``power-law'' singular value spectrum --- of problems encountered in ML for collaborative filtering.
Furthermore, the expanded data set reproduces some irregularities of the original data, in particular the accelerating decay of values in ranked row-wise and column-wise sums as well as in the singular values spectrum.

\subsubsection{Limitations}
\begin{figure}
    \centering
    \includegraphics[width=0.8\linewidth]{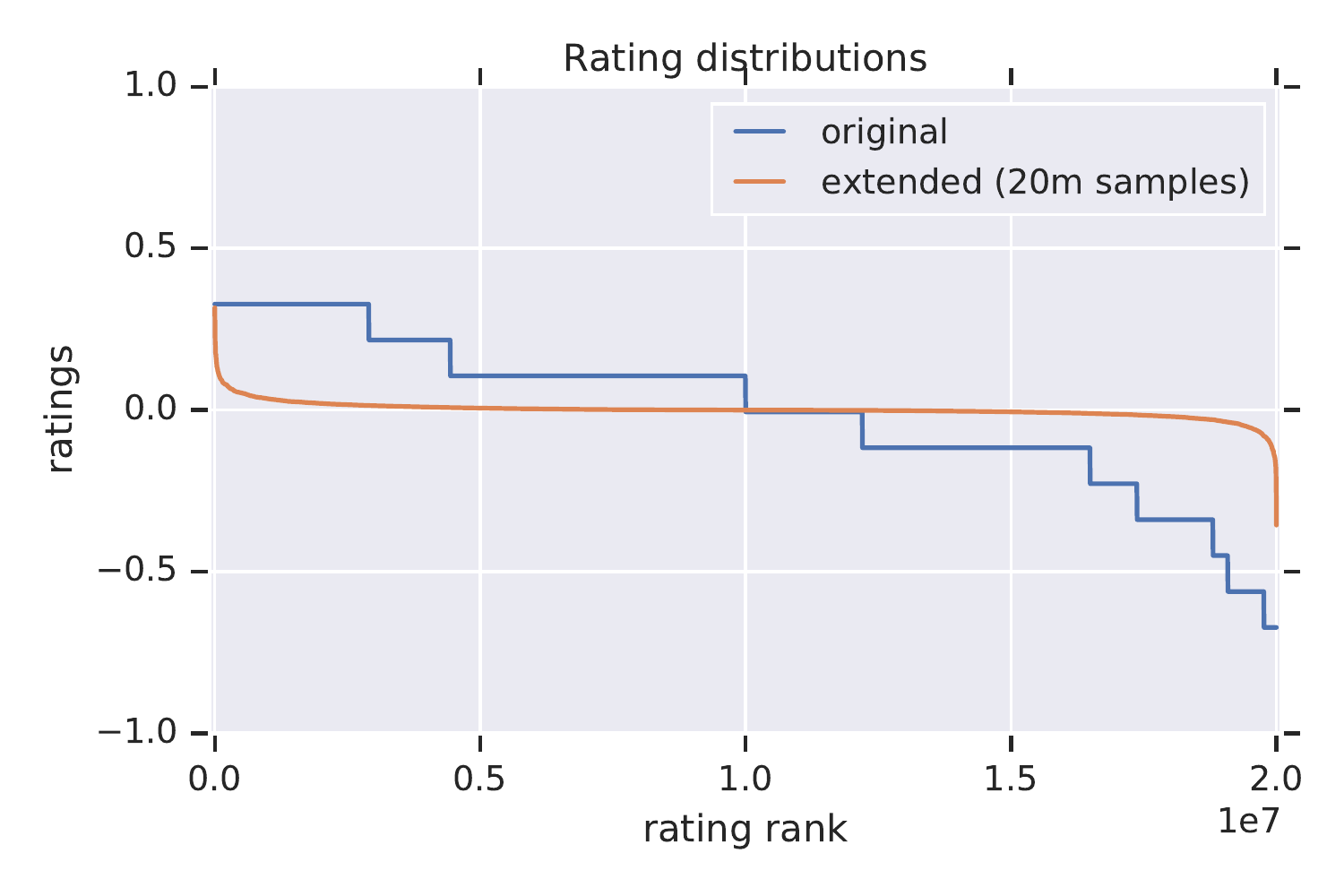}
    \caption{Sorted ratings in the original MovieLens 20m data set and the extended data. We sample 20m ratings from the new data set. Multiplications inherent to Kronecker products create a finer granularity rating scale which somewhat differs from the original rating scale. However, we can check that the new rating scale is still representative of common recommendation problems in that most rating values are close to the average and few user/item interactions are very positive or negative.}
    \label{fig:rating_distributions}
\end{figure}

Although it does not condition the difficulty of rating matrix factorization problems --- which depends primarily on the interaction distribution, the number of users, the number of items and sparsity of the rating matrix --- the distribution of ratings is still an important statistical property of the MovieLens 20m dataset. 
Figure~\ref{fig:rating_distributions} shows that there is a certain degree of divergence between the rating scales of the original and synthetic data set. 
In particular, many more rating values are present in the expanded data set as a result of the multiplication of terms from $\widehat{R}$ and $R$.
The transformations turning $R$ into $\widehat{R}$ create a smoother scale of values leading to a Kronecker product with many different possible ratings.
Although the non-zero value distribution is a divergence point between the two data sets, ratings in the synthetic data set are dominated by values which are close to the average  as in the original MovieLens 20m.
Therefore the synthetic ratings do have certain degree of realism as they represent user/item interactions where strong reactions (positive or negative) from users are much less likely than neutral interactions.

Another limitation of the synthetic data set is the block-wise repetitive structure of Kronecker products. 
Although the synthetic data set is still hard to factorize as the product of two low rank matrices because its singular values are still distributed similarly to the original data set, it is now easy to factorize with a Kronecker SVD~\cite{kamm2000optimal} which takes advantage of the block-wise repetitions in Eq~(\ref{eq:kronecker_product}).
Randomized fractal expansions which presented in Eq~(\ref{eq:gen_kron_product}) and Eq~(\ref{eq:random_sk_kron}) address this issue. A simple example of such a randomized variation around the Kronecker product consists in shuffling rows and columns of each block in Eq (\ref{eq:kronecker_product}) independently at random. The shuffles will break the block-wise repetitive structure and prevent Kronecker SVD from producing a trivial solution to the factorization problem.

As a result, the expansion technique we present appears as a reliable first candidate to train linear matrix factorization models~\cite{ricci2015recommender} and non-linear user/item similarity scoring models~\cite{he2017neural}.

%% file: conclusion.tex
In conclusion, this paper presents a first attempt at synthesizing a realistic large-scale recommendation data sets without having to make compromises in terms of user privacy.
We use a small size publicly available data set, MovieLens 20m, and expand it to orders of magnitude more users, items and observed ratings.
Our expansion model is rooted into the hierarchical structure of user/item interactions which naturally suggests a fractal extrapolation model.

We leverage Kronecker products as self-similar operators on user/item rating matrices that impact key properties of row-wise and column-wise sums as well as singular value spectra in an analytically tractable manner.
We modify the original Kronecker Graph generation method to enable an expansion of the original data by orders of magnitude that yields a synthetic data set matching industrial recommendation data sets in scale.
Our numerical experiments demonstrate the data set we create has key first and second order properties similar to those of the original MovieLens 20m rating matrix.

Our next steps consist in making large synthetic data sets publicly available although any researcher can readily use the techniques we presented to scale up any user/item interaction matrix.
Another possible direction is to adapt the present method to recommendation data sets featuring meta-data (e.g. timestamps, topics, device information).
The use of meta-data is indeed critical to solve the ``cold-start'' problem of users and items having no interaction history with the platform.
We also plan to benchmark the performance of well established baselines on the new large scale realistic synthetic data we produce.

%% file: appendix.tex
\subsection*{MovieLens 655 billion}
In this section, we present the properties of a larger expansion for MovieLens.
The reduced rating matrix $\widehat{R}$ is now of size $(128, 256)$ and the synthetic data consists of $655$ billion interactions between $17$ million users and $7$ million items.

\subsubsection*{Empirical properties of the reduced matrix $\widehat{R}$}
\begin{figure}
    \centering
    \includegraphics[width=0.85\linewidth]{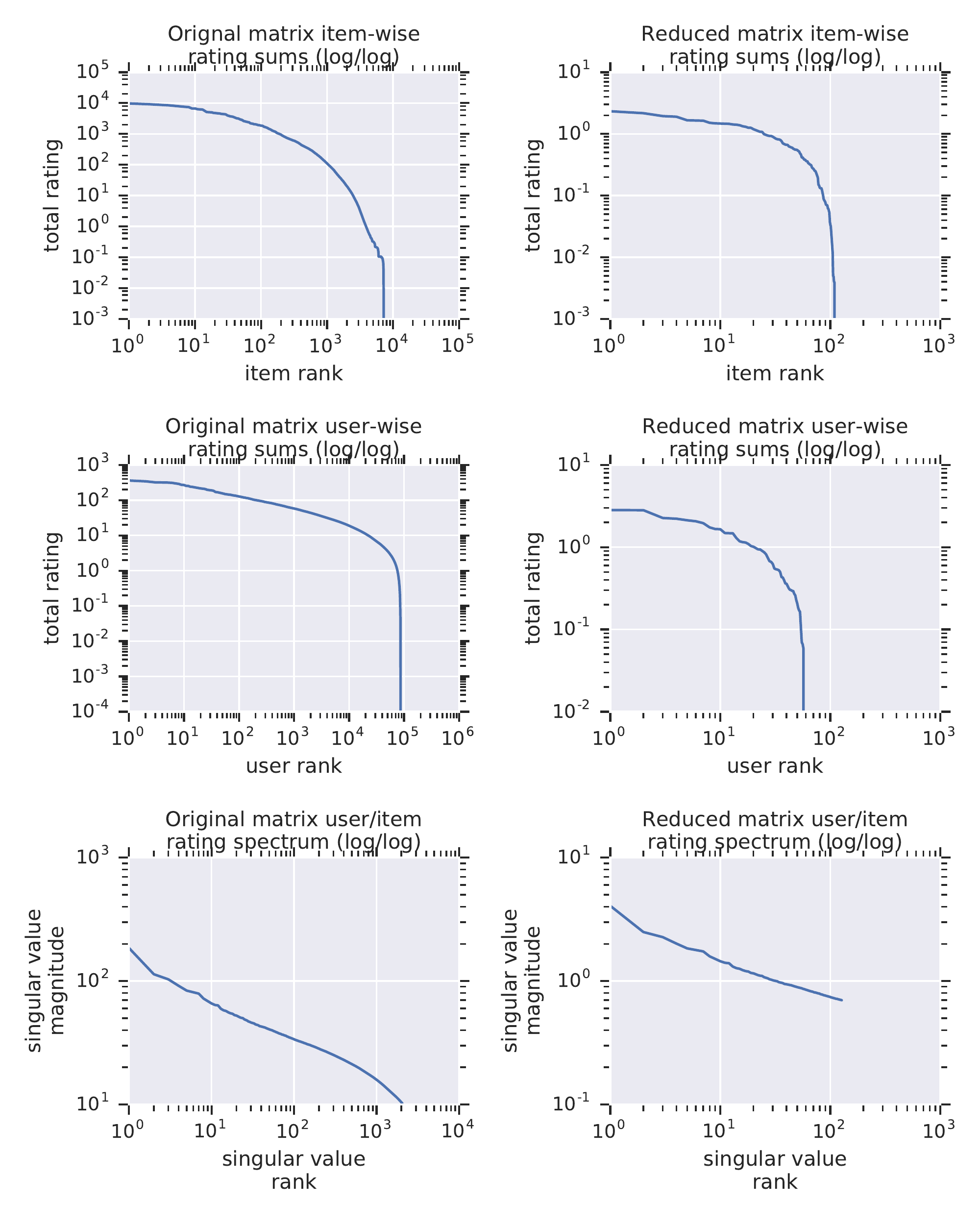}
    \caption{Properties of the reduced dataset $\widehat{R} \in \mathbb{R}^{128, 256}$. Once more, we validate the construction method numerically by checking that the distribution of row-wise sums, column-wise sums and singular values are similar between $R$ and $\widehat{R}$.}
    \label{fig:small_properties_XXL}
\end{figure}

We assess the scalability of the approach we present to synthesize $\widehat{R}$. In particular, we check that with a size of $(128, 256)$ instead of $(16, 32)$ $\widehat{R}$ still shares common statistical properties with the original matrix $R$.
Figure~\ref{fig:small_properties_XXL} demonstrates that the construction method we devised for $\widehat{R}$ still preserves key statistical properties of $R$.

\subsubsection*{Numerical validation for the expanded data set}
\begin{figure}
    \centering
    \includegraphics[width=0.85\linewidth]{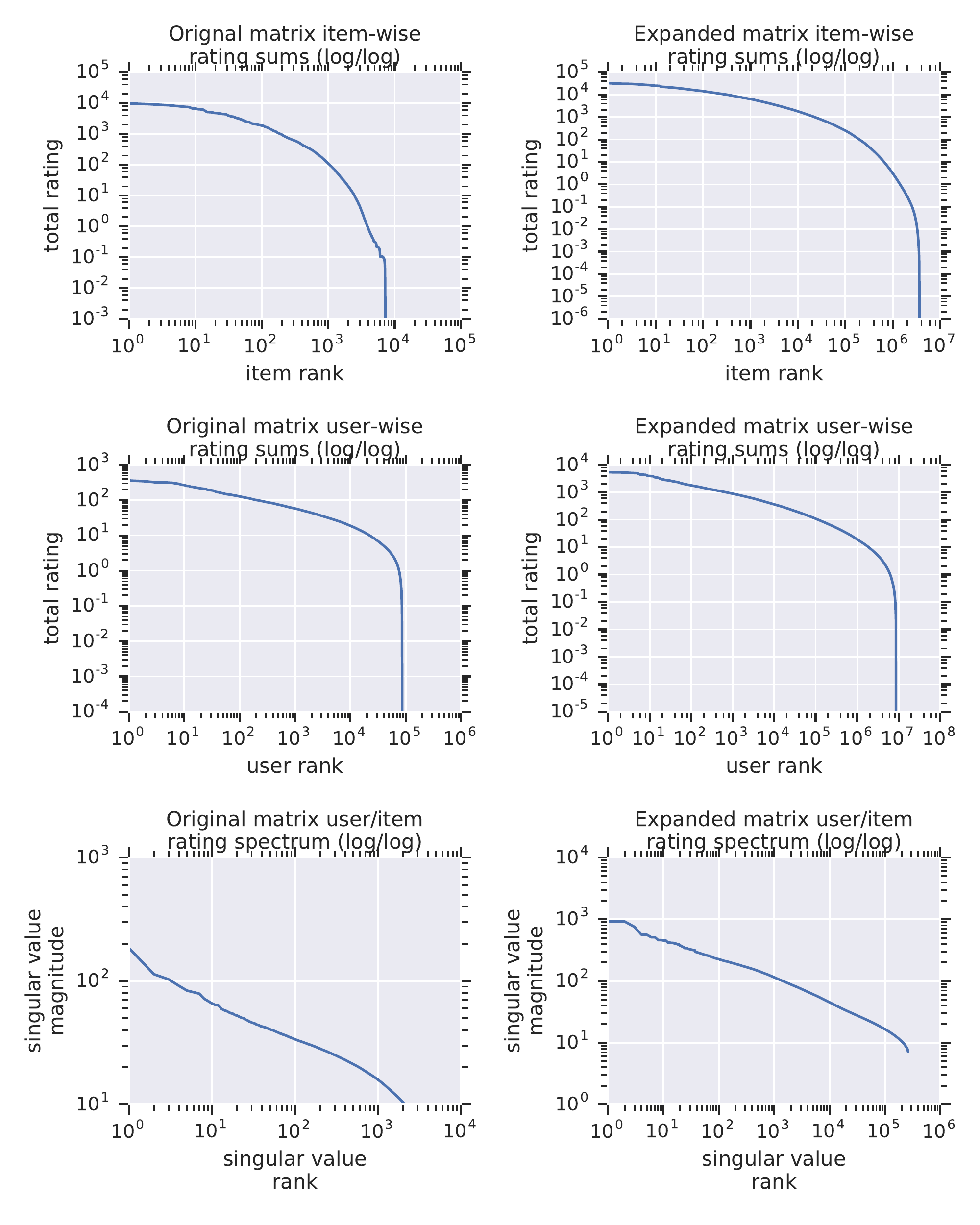}
    \caption{We again validate the construction method numerically by checking that the distributions of row-wise sums, column-wise sums and singular values are similar between $R$ and $\widehat{R} \otimes R$.
    Here as well, we can observe the similarity between key features of the original rating matrix and its synthetic expansion.
    }
    \label{fig:expanded_properties_XXL}
\end{figure}

We now verify that even with different extension factors and a much larger size, the synthetic data set we generate is similar to the original MovieLens 20m. We focus on the distribution of column-wise and row-wise sums in $\widetilde{R} = \widehat{R} \times R$ as well as the singular value distribution of the expanded matrix.
In Figure~\ref{fig:expanded_properties_XXL}, we find again that the ``power-law'' statistical behaviors and their inflections are preserved by the expansion procedure we designed.

\subsubsection*{Limitations}
\begin{figure}
    \centering
    \includegraphics[width=0.9\linewidth]{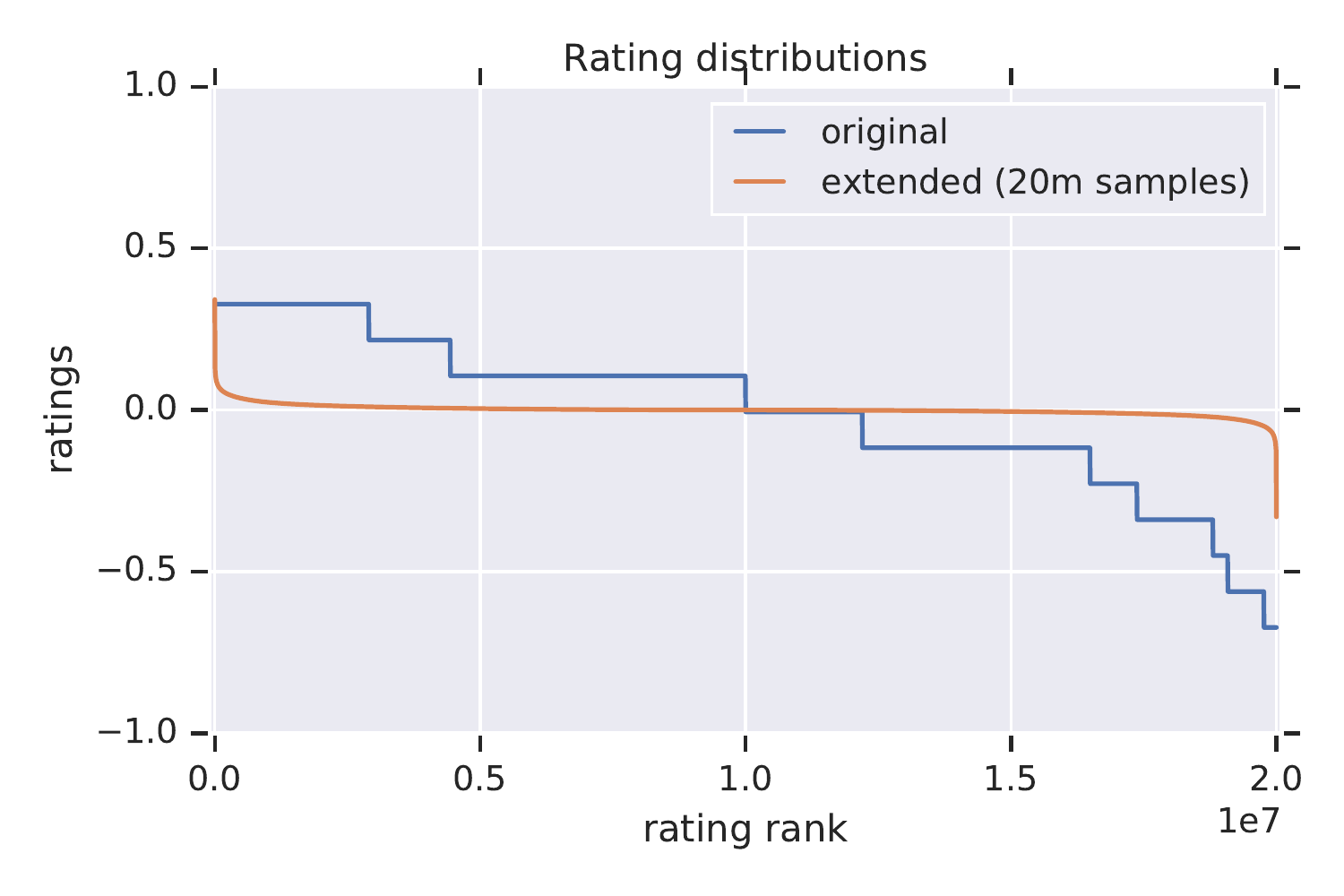}
    \caption{Sorted ratings in the original MovieLens 20m data set and the extended data. We sample 20m ratings from the new data set. Again we observe a certain divergence between the synthetic data set and the original. There is some realism though in that most interactions being near neutral.}
    \label{fig:rating_distributions_XXL}
\end{figure}

The previous observations demonstrate the scalability and robustness of our expansion method, even with an expansion factor of $(128, 256)$.
However, the same limitations are present as in the smaller case and Figure~\ref{fig:rating_distributions_XXL} shows that a similar divergence in rating scales exists between the original data set and its expanded synthetic version.
Like before the synthetic ratings remain realistic in that their majority is near average.
The present section therefore demonstrates that our method scales up and is able to synthesize very large realistic recommendation data sets.

